\documentclass[12pt]{article}
\usepackage{geometry}             
\usepackage{graphicx}
\usepackage{amssymb}
\usepackage{amsmath}
\usepackage{epstopdf}
\usepackage{comment}
\usepackage{cite}
\usepackage{caption}

\usepackage[hyperindex=true,
          pdfstartview=FitH,
          bookmarksnumbered=true,
          bookmarksopen=true,
          citecolor=blue,
          linkcolor=blue,
          colorlinks=true,
          pdfborder=001,
          unicode]{hyperref}

\allowdisplaybreaks

\begin{document}

\title{New class of rotating perfect fluid black holes in three dimensional gravity}
\author{Bin Wu$^{1}$ \ and Wei Xu$^{1,2}$\thanks{Correspondence author} \thanks{emails: {\it
binfen.wu@gmail.com} and {\it xuweifuture@mail.nankai.edu.cn}}\\
$^{1}$School of Physics, Nankai University, Tianjin 300071, China\\
$^{2}$School of Physics, Huazhong University of Science and Technology, \\ Wuhan 430074, China}
\date{}
\maketitle

\begin{abstract}
We obtain a new class of rotating black holes for Einstein theory with perfect
fluid source in (2+1) dimensions. We conclude that these black hole solutions only depend on variable angular velocity $m(r)$. Some examples of these black holes are given explicitly. In particular, the unknown static black hole in this special background  is obtained. In addition, the general properties including the horizon structure, energy conditions and equation of state, mass and angular momentum are explained in detail.
\end{abstract}

\section{Introduction}
Exact solutions of Einstein field equations have attracted considerable attentions since the General relativity is proposed, especially for cases with matter source. As the equations in the presence of matter are very complicated, to find exact solutions, a popular simplifying assumption has to be imposed that the matter field is a perfect fluid. We find that spacetime with perfect fluid source can be always considered as the interior spacetime of the black hole. The most famous example is schwarzchild interior solution. Furthermore, the spacetime with perfect fluid source is a good laboratory on testing some fundamental ideas which may apply to spacetime with other matter fields. The latest example is the application of holography in fluid dynamics, in which for any spherically symmetric black hole spacetime with a perfect fluid source, a dual hydrodynamics on a hypersurface near the black hole horizon can be established \cite{Wu:2013mda}. It is thus necessary to find the perfect fluid solutions, especially for the black hole solutions. Unfortunately, even with this specific assumption, very few solutions are found . In order to get rid of the difficulty, several papers investigated the situation when the spacetime is spherical symmetric(\cite{Lake:2002bq, Garcia:2002rn,Cataldo:2006yk, All-S, Spherical}). People also pay some attentions to 'static' cylindrically symmetry solutions(\cite{Cylinder1, Cylinder2}). The solutions in the modified gravity are shown in (\cite{Multamaki:2006ym, Brihaye:2011aa}), which correspond to $f(R)$ gravity and Massive gravity respectively. In this paper, we will focus on rotating perfect fluid black holes in three dimensional gravity. This is not actually a new idea. In fact four dimensional rotating perfect fluid solutions in axially-symmetric-cylindrical coordinates were well-known thirty years ago \cite{Rotating}. We aim at three dimensional gravity as it is often invoked in the study of AdS/CFT duality\cite{Brown:1986p547, Henneaux:2002p2538} and black hole physics. The general rotating solution for Einstein gravity with a perfect fluid source and cosmological constant in (2+1)-dimensions is obtained in \cite{Cataldo:2004fw}. Since we are interested in the three dimensional rotating perfect fluid black hole solutions, we will revisit this general rotating solution in a different gauge, based on which, we will show a new class of black holes and their properties.

The paper is organized as follows: we revisit the general three dimensional rotating perfect fluid solution in a different gauge in next section. In section 3, we obtain a new class of perfect fluid black hole solutions and their physical properties, including the geometric quantities, and unknown special degenerated static black hole. Mass and angular momentum of the black hole are also discussed. Finally we conclude by pointing out some interesting future tasks.

\section{General rotating perfect fluid solution in three dimensions revisited}
In this section, we revisit the general rotating perfect fluid solution \cite{Cataldo:2004fw} in a different gauge. In some of the publications in order to find perfect fluid solutions \cite{Garcia:2002rn, All-S, Multamaki:2006ym}, the Oppenheimer-Volkov equation \cite{Oppenheimer:1939ne}, i.e., the energy-momentum conservation, has been used. While we prefer solving the Einstein field equation directly in our case, since the energy-momentum is automatically conserved. We begin with the Einstein equations with a source term
\begin{align}
&  G_{\mu\nu}=R_{\mu\nu}-\frac{1}{2}Rg_{\mu\nu}=-\Lambda g_{\mu\nu} + \kappa T_{\mu\nu}, \nonumber
\end{align}
where $G_{\mu\nu}$ is the Einstein tensor, $T_{\mu\nu}$ is the stress-energy-momentum tensor and $\Lambda=-\frac{1}{\ell^2}$ is the (bare) cosmological constant. Here, we force the cosmological constant to be negative, for the reason that smooth black hole horizons in three dimensions can exist only in the presence of a negative cosmological constant\cite{Ida:2000jh}. In our following discussion, we will set the gravitational constant $\kappa=1$ for convenience.

We are interested in the perfect fluid solutions, the stress-energy-momentum tensor can be written as
\begin{align*}
  T_{\mu\nu}=(\rho(r)+p(r))u_{\mu}u_{\nu}+p(r) g_{\mu\nu},
\end{align*}
where $u_{\mu}$ is the proper velocity. We also assume that both the the energy density $\rho(r)$ and fluid pressure $p(r)$ depend only on the radial coordinate $r$. One can further assume that there exist a relation between $\rho$ and $p$,
\begin{align}
f(\rho,p)=0,
\label{EOS}
\end{align}
called {\it equation of state}.

To obtain rotating solution, we begin with the following general metric ansatz
\begin{align}
  \mathrm{d} s^2=-f(r)\mathrm{d} t^2+\frac{1}{g(r)}
  \mathrm{d} r^2+r^2\bigg(\mathrm{d} \theta-m(r)\mathrm{d} t\bigg)^2,
  \label{lineelm}
\end{align}
where the coordinate ranges are given by $-\infty<t<\infty$, $r\geq0$,
$-\pi\leq\theta\leq\pi$ and $m(r)$ is the angular velocity. Firstly, we calculate a few lines to get the proper velocity of the perfect fluid in above spacetime. If we set $x^\mu=(x^0,x^1,x^2)=(t,r,\theta)$, the components of the proper velocity are
\begin{align*}
u^{\mu}=\frac{d\,x^{\mu}}{d\,\tau}=\left({\frac {1}{\sqrt {f \left( r \right) }}}, 0, {\frac {m(r)}{\sqrt {f \left( r \right) }}}  \right).
\end{align*}
We can also find it through using the Killing vector field $K^\mu$, which is proportional to the proper velocity $u^{\mu}$, i.e. $u^\mu = \frac{1}{V}K^\mu$. Here $K^\mu=(1, 0, m(r))$ is a linear combination of killing vector field $(\partial_t)^\mu$ and rotational killing vector field $(\partial_\theta)^\mu$, and $V= \sqrt{-K^{\mu} K_{\mu}}=\frac{1}{\sqrt {f(r)}}$ is the red-shift factor.

Given the definition for the equation array
\begin{align}
Eq^\mu{}_\nu\equiv G^\mu{}_\nu-T^\mu{}_\nu-\frac{1}{\ell^2} g^\mu{}_\nu=0,
\label{eineq}
\end{align}
the Einstein equations for metric (\ref{lineelm}) can be found explicitly and listed in Appendix A because of their complicated forms. In addition, there are continuity equation and conserved equations for energy-momentum
stress tensor of the perfect fluid, i.e.
\begin{align*}
 \nabla_\mu(\rho u^\mu)&=0,   \\
 \nabla_\mu T^{\mu\nu} &=0.
\end{align*}
In our case, the continuity equation is always satisfied, and only the $r$-component
of the conserved equations is non-vanishing, as is shown below
\begin{align}
g \left( \rho\,f_{{r}}+pf_{{r}}+2\,p_{{r}}f \right) =0,   \label{conserved}
\end{align}
where $p_r$ and $f_r$ represent the first derivative respect with r of $p(r)$ and $f(r)$.

Solving the $Eq^t{}_\theta$ component of Eq.(\ref{eineq}), i.e. Eq.(\ref{ein2}), one arrives at
\begin{align}
 g={\frac {{\it c_0}\,}{{m_{{r}}}^{2}{r}^{6}}} \,f ,
 \label{gg}
\end{align}
where $c_0$ is an integration constant. Without loss of generality, we can assume $c_0=1$ in the rest of the paper. The other choices that $c_0$ is arbitrary constant can be recovered by
rescaling the $t$ coordinate, i.e. $f\to \frac{f}{c_0}, t\to c_0t$.

Then taking the combination of $Eq^t{}_t$, $Eq^r{}_r$
and $Eq^\theta{}_t$ components of Eq.(\ref{eineq}), i.e. Eq.(\ref{ein1}, \ref{ein3}, \ref{ein4}), one obtains another relationship about the function $f(r),g(r)$ and $m(r)$. This relationship together with Eq.(\ref{gg}) give rise to
\begin{align*}
 2\,{r}^{3}{m_{{r}}}^{3}+4\,m_{{r}}f
_{{r}}+rm_{{r,r}}f_{{r}}-f_{{r,r}}rm_{{r}} =0,
\end{align*}
which yields
\begin{align}
  f=\int \! \left( 2\,\int \!{\frac {m_{{r}}}{{r}^{2}}}{dr}+{\it c_1}
\right) {r}^{4}m_{{r}}{dr}+{\it c_2},
\label{ff}
\end{align}
with $c_1$ and $c_2$ being integration constants.

Finally, the other components of Eq.(\ref{eineq}) and Eq.(\ref{gg}) are used to derive the energy density $\rho(r)$ and pressure $p(r)$
\begin{align}
&p(r)=\Lambda+\frac{{r}^{3}{m_{{r}}}^{2}g+2\,f_{{r}}g}{4rf},  \label{pp}\\
& \rho(r)= -\Lambda+\frac {4\,grfm_{{r,r}}+12
\,gfm_{{r}}-{r}^{4}{m_{{r}}}^{3}g-2\,grm_{{r}}f_{{r}}}{4{r}^{2}fm_{{r}}},
\label{rhorho}
\end{align}
which show that the conserved equation Eq.(\ref{conserved}) is satisfied automatically.

Then we get the general rotating perfect fluid solution \cite{Cataldo:2004fw} in a different gauge, where the metric is (\ref{lineelm}), $f(r)$ behaves as Eq.(\ref{ff}), $g(r)$  behaves as Eq.(\ref{gg}) with an arbitrary angular velocity $m(r)$, and the perfect fluid source is characterized by the pressure $p(r)$ given in Eq.(\ref{pp}), density $\rho(r)$ shown in Eq.(\ref{rhorho}). One need note that the uncertainly of angular velocity $m(r)$ is bare, actually the uncertainly of solution is emerged from the density $\rho(r)$, namely from the matter source. Hence this uncertainly is behoove appears,
if no extra constraint condition is imposed. The same phenomenon also happens in
the study of all three dimensional static circularly symmetric perfect fluid solutions \cite{Garcia:2002rn}. However, for physically conceivable  solutions, one of the energy conditions for pressure $p(r)$ and density $\rho(r)$ must be satisfied at least. Simultaneously, the choice of $\rho(r)$ is restricted by physically reasonable matter distributions.

Now let us go back to the function $f(r)$. If we consider a vanishing perfect fluid matter distribution, namely the $p(r)=0,\rho(r)=0$ limit of the above general solution, one find that $c_1$ and $c_2$ are related to cosmological constant $\Lambda=-\frac{1}{\ell^2}$ and the black hole mass $M$ via
\begin{align*}
  c_1=-\frac{a}{\ell^2},\quad c_2=-M,
\end{align*}
in which case the solution degenerates into three dimensional black hole in the pure gravity with a negative cosmological constant, i.e. the BTZ black hole. Here $a$ is the rotating factor in the angular velocity $m(r)=\frac{a}{r^2}$ of BTZ black hole \cite{Banados:1992wn}. It is similar to $c_0$ that we can choose $c_1=-\frac{1}{\ell^2}=\Lambda$, while the other choices can be recovered by rescaling the $\theta$ coordinate, i.e. $\theta\to a\theta$.

\section{New class of three dimensional rotating perfect fluid black hole solutions}
In this section, we aim to obtain new rotating perfect fluid black hole. It is worth mentioning that the general rotating perfect fluid solutions characterized by Eq.(\ref{gg},\ref{ff},\ref{pp},\ref{rhorho}) with variable density $\rho(r)$ have only two parameters, i.e. $c_1$ and $c_2$ from $f(r)$. In order to get some concrete and new exact black holes with all the function being analytic expressions, especially for $f(r)$, one can begin with a concrete fluid matter with a known $\rho(r)$. However, this is not an usual way, since a concrete and physically reasonable fluid is always interpreted as matter fields, such as Maxwell field \cite{Martinez:1999qi,Martinez:2006an}, scalar field \cite{Henneaux:2002wm,Martinez:1996gn,Martinez:2004nb,Xu:2013nia,Zhao:2013isa}, higher rank tensor fields \cite{Perez:2013xi,Ammon:2011nk,Chen:2012pc} and higher curvature terms \cite{Oliva:2012ff,Bagchi:2011vr,Li:2008dq,Bergshoeff:2009hq}. However, throughout the paper, we focus our attention on the pure fluid matter rather than known ones. That is totally a different method compared with the conventional studies.

As shown in the above section, the solution only depends on one arbitrary function, which calls for an extra constraint in the solution. In \cite{Cataldo:2004fw}, people assume that the perfect fluid rotating solutions have known equation of state (i.e. Eq.(\ref{EOS})), such as the linear law $p=\omega \rho$ and polytropic law $p=C\rho^{\gamma}$. This is the most popular extra input to find perfect fluid solution. Again, BTZ black hole can be obtained as a special branch of the case of linear law, which is out of our interest. Actually, one does not need to add such an extra constraint in our case, because there is a natural and intrinsic constraint on angular velocity $m(r)$, i.e. whose asymptotic behavior at $r\rightarrow+\infty$ behaves as
\begin{align}
m(r)|_{r\rightarrow+\infty}=0,
\label{asy}
\end{align}
after removing the global rotations of the coordinate system. One can naturally propose a finite-polynomial solution
\begin{align}
  m_{n}(r)=a\,\sum_{i=2}^{n} a_{i} \, r^{-i},
  \label{angular}
\end{align}
where $-n$ is the lowest power of $r$ in $m_n(r)$, and the highest power must be $-2$, in order to include the well known BTZ black hole as a simplified limit of (\ref{angular}). After inserting the angular velocity, one can get $f(r)$ from Eq.(\ref{ff}), $g(r)$ from Eq.(\ref{gg}), $p(r)$ from Eq.(\ref{pp}) and $\rho(r)$ from Eq.(\ref{rhorho}), and all the five functions make up a new exact perfect fluid solution, which is a black hole as will be shown later. The same phenomenon that derives rotating hairy solutions with an infinite asymptotic behavior as Eq.(\ref{asy}) happens in \cite{Zhao:2013isa}, whose angular velocity belongs to the subcase with $n=3$ appearing in Eq.(\ref{angular}). This method is different from that in \cite{Cataldo:2004fw}. The interesting point here is that, the equation of state is a derived object rather than an extra input, once the Eq.(\ref{angular}) is imposed. Therefore one can say that this new class of rotating perfect fluid black hole solutions are more ``general" and they depend on the angular velocity $m(r)$ with an infinite asymptotic behavior as Eq.(\ref{asy}).

However, one can expect that the other functions ($f(r),g(r),p(r),\rho(r)$) of this new class of black hole have complicated forms, which makes people unable to explore their properties, especially for the perfect fluid source. Hence, in next subsection, we will show a simple example with $n=3$ of this class of black hole, which is the next order of the simplest one (the BTZ black hole for $n=2$). Based on this example, we will give overview of its properties. One can also generalize this study to the general class of the black holes by adapting the same procedure.

 Now we focus on the case with $n=3$, i.e. $m_{3}(r)=a(a_2r^{-2}+a_3r^{-3})$. This kind of angular velocity is different from the case in  the pure three dimensional gravity (the case for $n=2$). We can find a similar case in \cite{Zhao:2013isa}, which shows a rotating hairy black hole. The parameter $a_2$ is related to the pure gravity, and $a_3$ is related to the scalar field. Without loss of generality, we choose the same angular velocity
\begin{align}
  m(r)={\frac {a \left( 3\,r+2\,B \right) }{{r}^{3}}},
  \label{solm}
\end{align}
in order to find physical meaning of the parameters in our case comparing with \cite{Zhao:2013isa}. Here $a$ is chosen to be positive, while the negative side can be recovered by the coordinate transformation of $\theta$, i.e. $\theta\rightarrow-\theta$. Thus Eq.(\ref{gg},\ref{ff},\ref{pp},\ref{rhorho}) show the other  structural functions of this solution as follows,
\begin{align}
  &f(r)= {\frac {{r}^{2}}{{\ell}^{
2}}}+\,{\frac {2Br}{{\ell}^{2}}}-M+9{a}^{2}\left( \frac{1}{r^2}+\frac{6B}{5r^3}+\frac{2B^2}{5r^4}\right) \label{solf},\\
  &g(r)=\frac{r^2}{(r+B)^2}f(r)\label{solg},\\
  &p(r)=\,{\frac {B \left( 9\,{a}^{2}{\ell}^{2}-5\,{r}^{4} \right) }{5{\ell}^{2}{r
}^{4} \left( r+B \right) }}\label{solp},\\
  &\rho(r)={\frac {B \left( {B}^{2}+M{\ell}^{2} \right) }{ \left( r+B \right) ^{3}
\,{\ell}^{2}}}-\,{\frac {9B{a}^{2} \left( 6\,{r}^{2}+8\,Br+3\,{B}^{2} \right) }{
5\,{r}^{4} \left( r+B \right) ^{3}}},\label{solrho}
\end{align}
where, the constant $M$ and $a$ are associated with the conserved charges mass and angular momentum respectively, and $B$ characterizes the property of perfect fluid source in some sense.  Eq.(\ref{lineelm},\ref{solm},\ref{solf},\ref{solg},\ref{solp},\ref{solrho}) constitute a full set of an exact new three dimensional perfect fluid solution. In next subsection, we will mainly focus our attention  on the physical properties of this solution to have a further understanding.

\subsection{Geometric quantities}
As we are interested in three dimensional black hole solutions, we need to calculate some
of the associated geometric quantities to further characterize the geometry of the solutions. First of all, the Ricci scalar \begin{align}
  R=\,{\frac {2BM}{ \left( r+B \right) ^{3}}}-\,{\frac {18B{a}^{2}
 \left( 8\,{r}^{2}+12\,Br+5\,{B}^{2} \right) }{5{r}^{4} \left( r
+B \right) ^{3}}}-\,{\frac {2r \left( 3\,{r}^{2}+7\,Br+5\,{B}^{2}
 \right) }{ \left( r+B \right) ^{3}{\ell}^{2}}},
 \label{Ricci}
\end{align}
which corresponds to two curvature singularities at $r=0$ and $r=-B$ if $a\neq0,B\neq0$, while it has only one at $r=-B$ if $a=0,B\neq0$ and no singularity in the case $B=0$. The singularities of higher order curvature invariants such as $R_{\mu\nu}R^{\mu\nu}$ and
$R_{\mu\nu\rho\sigma}R^{\mu\nu\rho\sigma}$ have a similar behavior. We can conclude that the corresponding solution is a black hole solution if there is a singularity with a round horizon. The classification of singularities gives a clue for searching for some special cases of this perfect fluid black hole solution.

The Cotton tensor is defined as
\begin{align*}
C_{\mu\nu\sigma}=\nabla_\sigma R_{\mu\nu}-\nabla_\nu R_{\mu\sigma}
+\frac{1}{4}(\nabla_\nu R g_{\mu\sigma} - \nabla_\sigma R g_{\mu\nu}).
\end{align*}
For our solution, there are some non-vanishing components of cotton tensor, we only list a simple one below:
\begin{align*}
 C_{\theta\theta r}& =\,{\frac { 27\left( 10\,{r}^{3}+20\,B{r}^{2}+15\,r{B}^{
2}+4\,{B}^{3} \right) B{a}^{2}}{10{r}^{3} \left( r+B \right) ^{4}}}-\,
{\frac {3BM{r}^{2}}{ 2\left( r+B \right) ^{4}}}-\,{\frac {3{B}^{3}{r}^
{2}}{2 \left( r+B \right) ^{4}{\ell}^{2}}}.
\end{align*}
When $B\neq0$, the non-vanishing Cotton tensor signifies that the metric is not
conformally flat \cite{Cotton}.

\subsection{Unknown static degenerated cases}
When $a=0$, the solution degenerates into an unknown static perfect fluid black hole, which reads as
\begin{align*}
  &f(r)={\frac {{r}^{2}}{{\ell}^{2}}}+\,{\frac {2Br}{{\ell}^{2}}}-M,\\
  &g(r)=\frac{r^2}{(r+B)^2}f(r),\\
  &p(r)=-{\frac {B}{ \left( r+B \right) {\ell}^{2}}},\\
  &\rho(r)={\frac {B \left( M{\ell}^{2}+{B}^{2} \right) }{{\ell}^{2} \left( r+B
 \right) ^{3}}},
\end{align*}
and the Ricci scalar Eq.(\ref{Ricci}) degenerates to
\begin{align*}
  R=\,{\frac {2BM}{ \left( r+B \right) ^{3}}}-\,{\frac {2r \left( 3\,{r}^{
2}+7\,Br+5\,{B}^{2} \right) }{ \left( r+B \right) ^{3}{\ell}^{2}}},
\end{align*}
which shows a curvature singularity located in $r=-B$. The possible range of radial coordinate $r$ is thus corrected to $r>-B$ according to the existence of singularity.\footnote{One can also choose the $r<-B$ side.} The black hole has a horizon located in
\begin{align*}
r_0=-B+\sqrt {{B}^{2}+M{\ell}^{2}},
 \end{align*}
while another solution of $f(r)=0$, i.e. $r=-B-\sqrt {{B}^{2}+M{\ell}^{2}}$ is out of the radial coordinate ranges and is discarded. We find that the existence of black hole horizons imposes an lower bound for the mass parameter $M$, which reads
\begin{align}
  M>-\frac{B^2}{\ell^2}.
  \label{MM}
\end{align}
When $M=-\frac{B^2}{\ell^2}$, the horizon is located in $r=-B$ and the curvature singularity at $r =-B$ will be naked, which is not physically interesting. One can see the pressure $p(r)$ and density $\rho(r)$ are also singular in $r=-B$. Furthermore, one can easily find these relations
 \begin{align}
   \rho p<0, \quad\rho+p=\frac{-B}{(r+B)^3}f(r),  \label{ec}
 \end{align}
which show that when $B>0$, all energy conditions fail. When $B<0$, the null energy condition (NEC) and strong energy condition (SEC)
 \begin{align}
   \quad\rho+p>0,\quad \rho+3p>0,
   \label{eos11}
 \end{align}
hold in the static region of the spacetime, i.e. $r>r_0$. However, the weak and dominant energy conditions still fail as $\rho>0$ and $\rho>|p|$. For the general rotating perfect fluid black hole characterized by Eq.(\ref{solm},\ref{solf},\ref{solg},\ref{solp},\ref{solrho}), these two equation (\ref{ec},\ref{eos11}) always hold. That to say, in order to satisfy the null energy condition (NEC) and strong energy condition (SEC) in the static region, $B<0$ should be necessary (it works with $B>0$ if we choose the radial coordinate range as $r<-B$). Therefore, in the following discussion, we always choose $B<0$.

Finally the parameters in the fluid can be further reduced when we solve the $\rho(r)$ and $p(r)$ simultaneously,
\begin{align}
  \rho+{\frac {{\ell}^{4} \left( {B}^{2}+M{\ell}^{2} \right) }{{B}^{2}}}{p}^{3}=0.
  \label{EOS1}
\end{align}
The Eq.(\ref{EOS1}) is exactly the same as the polytropic law $p=C\rho^{\gamma}$ with $\gamma=\frac{1}{3}$. Therefore, we prove the fact that the equation of state is a derived object. Here, the parameter $B$ characterizes the equation of state of the perfect fluid source. Though this equation of state shows the negative energy density, Eq.({\ref{eos11}}) tell us that it does not violates all reasonable energy condition, as the null energy condition (NEC) and strong energy condition (SEC) are satisfied in the static region of the spacetime. Back to the rotating black hole, one can follow the same procedure to find its equation of state in principle, which is a little complicated.

\subsection{Mass and angular momentum}
In this subsection, we present the formulas for calculating the mass and angular momentum shown by Brown-York \cite{Brown:1992br,Brown:1994gs,Creighton:1995au}. Using this formalism, one can obtain the quasi-local energy
$G(r)$ at a radial boundary $r$
\begin{align*}
G(r) = 2\bigg(\sqrt{g_0(r)}-\sqrt{g(r)}\bigg),
\end{align*}
and the quasi-local angular momentum $j(r)$
\begin{align*}
j(r) =-r^3\sqrt{ \frac{g(r)}{f(r)}}\frac{\partial m(r)}{\partial r},
\end{align*}
In our case, $g_0(r)=\frac{r^2}{\ell^2}$ is the background metric function. And the quasi-local mass
$F(r)$ is given by
\begin{align*}
F(r) = G(r)\sqrt{f(r)} - j(r) m(r).
\end{align*}
Then the mass and angular momentum
are defined and calculated to be
\begin{align*}
  &E\equiv F(+\infty)=M+\frac{B^2}{\ell^2},\\
  &J\equiv j(+\infty)=6a,
\end{align*}
respectively. One can note that $E>0$ is consistent with the horizon condition for degenerated static black hole, i.e Eq.(\ref{MM}).
\section{Conclusion}
In this paper, we obtain a new class of rotating black holes for Einstein theory with a perfect
fluid source in (2+1) dimensions without extra constraints, such as equation of state. In the sense, one can say that this new class of rotating perfect fluid black hole solutions are more ``general" and depend on variable angular velocity $m(r)$, which is a finit-polynomial solution resulted from a vanishing infinite asymptotic behavior as shown in Eq.(\ref{asy}). Some examples of these black holes are shown mainly. Their physical properties are presented as well, such as the geometric quantities. The corresponding degenerated unknown static black hole is shown with its horizon structure, energy conditions and equation of state. Especially for the latter two, one can find that the parameter $B$ appearing in the angular velocity $m(r)$ is negative (in the $r>-B$ side) and it characterizes the equation of state of the perfect fluid source. Mass and angular momentum of the black hole are also discussed.

For different angular velocity $m_n(r)$, one can follow the same procedure to get new rotating black holes. Meanwhile, corresponding new static black holes can be obtained as the vanishing angular momentum limit of the rotating ones. Note the physical solutions can be chosen by the energy conditions. One can expect there are interesting properties of these black holes for their complicated forms. Usually people use the energy-momentum tensor directly to solving the Einstein equation in the presence of known matter field source. In this sense, we consider this method as a new way to search the black hole solutions in the presence of unknown matter field source. There remain some other interesting problems for further studying:
the physical interpretation for the perfect fluid source and thermodynamics for the perfect fluid solutions \cite{Coll:1996ux,Quevedo:1994yg,Coll:2004jm,Coll:2003tv}. An effective way is their Lagrange and Hamiltonian description \cite{Krasinski:1997zz,Coll:2000uy,Brown:1992kc}.

\section*{Acknowledgements}
We would like to thank Liu Zhao and Bin Zhu for useful conversations. Bin Wu is supported by the Ph.D. Candidate Research Innovation Fund of Nankai University.

\section*{Appendix A}
The Einstein equations (\ref{lineelm}) with cosmological constant and
perfect fluid source for metric (\ref{eineq})  is explicitly listed below:
\begin{align}
Eq^t{}_t&=
\frac{1}{4f^2 r} \bigg(g{r}^{3}f{m_{{r}}}^{2}+g_{{r}}{r}^{3}fmm_{{r}}
+6\,g{r}^{2}fmm_{{r}}-g{r}^{3}f_{{r}}mm_{{r}}  \nonumber  \\
&\quad+2\,g{r}^{3}fmm_{{r,r}} +2\,g_{{r}}{f}^{2}+4\,{f}^{2}\rho\,r+4\,\Lambda\,r{f}^{2}\bigg),   \label{ein1}\\
Eq^t{}_\theta& =\frac{r}{4f^2}\bigg( grm_{{r}}f_{{r}}-g_{{r}}rfm_{{r}}-6\,gfm_{{r}}-2\,grf
m_{r,r}\bigg), \label{ein2} \\
Eq^r{}_r&= \frac{1}{4f^2} \bigg( {r}^{3}{m_{{r}}}^{2}g+2\,f_{{r}}g-4\,prf+4\,\Lambda\,rf  \bigg),\label{ein3}  \\
Eq^\theta{}_t &= -\frac{1}{4f^2 r}\bigg( g{r}^{3}f_{{r}}{m}^{2}m_{{r}}
+mg_{{r}}rff_{{r}}+2\,mgrff_{{r,r}}-mgr{f_{{r}}}^{2}+fgrm_{{r}}f_{{r}}  \nonumber \\
&\quad-2\,g{r}^{3}f{m}^{2}m_{{r,r}}-6\,g{r}^{2}f{m}^{2}m_{{r}}-g_{{r}}{r}^{3}f{m}^{2}m_{{r}}
-4\,fg{r}^{3}{m_{{r}}}^{2}m   \nonumber \\
&\quad-g_{{r}}r{f}^{2}m_{{r}}-2\,g_{{r}}{f}^{2}m-6\,g{f}^{2}m_{{r}}-4\,m{
f}^{2}\rho\,r-2\,gr{f}^{2}m_{{r,r}}-4\,prm{f}^{2}\bigg),   \label{ein4}\\
Eq^\theta{}_\theta&= -\frac{1}{4f^2}\bigg( {r}^{2}mg_{{r}}fm_{{r}}-{r}^{2}mgm_{{r}}f_{{r}}+6\,rmgfm_{{r}}+2\,{r}^{2}mgfm_{{r,r}} \nonumber \\
&\quad+3\,g{r}^{2}f{m_{{r}}}^{2}+g{f_{{r}}}^{2}-2\,gff_{{r,r}}-g_{{r}}ff_{{r}}+4\,p{f}^{2}-4\,\Lambda\,{f}^{2}\bigg)\label{ein5},
\end{align}
where $m_r, p_r$ and $f_r$ represent the first derivative for r, $m_{r,r}$ is second derivative and so on.

\providecommand{\href}[2]{#2}\begingroup
\footnotesize\itemsep=0pt
\providecommand{\eprint}[2][]{\href{http://arxiv.org/abs/#2}{arXiv:#2}}

\end{document}